\def\sqig{$\sim\,$} 
\def\etal{et\,al.}
\def\sqiglt{\hbox{\rlap{\lower.55ex \hbox {$\sim$}}
        \kern-.3em \raise.4ex \hbox{$<$}\,}}
\def\pten#1{$\times10^{#1}$}
\def\appro{$\approx\,$}

\def\he2{He\,{\sc ii}\,$\lambda$1640}
\def\heii{He\,{\sc ii}}
\def\c4{C\,{\sc iv}\,$\lambda$1550}
\def\civ{C\,{\sc iv}}
\def\kmps{km\,s$^{-1}$}
\documentclass{emulateapj}
\shorttitle{The magnetic accretor AO~Psc}
\shortauthors{Hellier \&\ van Zyl}
\begin{document}
\title{Stream--field interactions in the magnetic accretor AO~Piscium}
\author{Coel Hellier and Liza van Zyl}
\affil{Astrophysics Group, Keele University, Staffordshire, ST5 5BG, U.K.}
\begin{abstract}
UV spectra of the magnetic accretor AO~Psc show absorption features
for half the binary orbit. The absorption is unlike the wind-formed
features often seen in similar stars. Instead, we attribute it to a
fraction of the stream that overflows the impact with the accretion
disk.  Rapid velocity variations can be explained by changes in the
trajectory of the stream depending on the orientation of
the white-dwarf's magnetic field.  Hence we are directly observing the
interaction of an accretion stream with a rotating field. We compare
this behavior to that seen in other intermediate polars and in SW~Sex
stars.
\end{abstract}
\keywords{accretion, accretion disks --- stars: 
individual(\objectname{AO~Psc}) --- novae, cataclysmic variables --- 
binaries: close}

\section{Introduction}
Cataclysmic variable stars (see Warner 1995) give us an opportunity
for studying accretion under a range of physical
conditions. For example, the AM Her class shows an accretion stream
interacting with a magnetic dipole that is stationary with respect to
the stream.  In the intermediate polars (IPs) the white dwarf is generally 
spinning much faster than the orbit, and the stream material usually 
circularises into an accretion disk before threading onto the field.
However, there are signs that at least part of the
stream often overflows the accretion disk and interacts directly with the
magnetic field (e.g.\ Hellier \etal\ 1989) giving a 
more dynamic and poorly understood stream--field interaction. In at least
one system, V2400~Oph, stream-fed accretion appears to dominate 
(Buckley \etal\ 1995; Hellier \&\ Beardmore 2002). 

AO~Psc is an IP with a 3.59-hr orbit and an 805-s white-dwarf spin
period whose optical and X-ray behavior have been well studied
(e.g.\ Hellier, Cropper \&\ Mason 1991, hereafter H91; Hellier \etal\ 1996;
Taylor \etal\ 1997; Williams 2003).  We report here on UV spectroscopy
of AO~Psc which shows a new type of observational evidence for a
stream--field interaction, namely rapidly varying absorption features
in UV metal lines.

\section{Observations and spectra}
On 2000 July 15 and 16 we obtained UV observations of AO Psc with
{\sl HST}'s STIS spectrograph. The E140M echelle grating gave 43
orders covering the wavelength range 1140--1710~\AA, with a dispersion
of 0.015~\AA\ at the 1425-\AA\ central wavelength. The observations
were made with the FUV-MAMA detector in time-tag mode, obtaining data
trains lasting 2000--3000 s from each of eight {\sl HST\/} orbits.

We used the STSDAS package {\sc inttag} in {\sc iraf} to bin the data
train into 40-s exposures. This gave 474 echelle images which were
then reduced to spectra using the STSDAS spectral reduction package
{\sc calstis}.

The summed spectrum is shown in Fig.~1.  The 
lines are predominantly in emission, which is unusual among UV
spectra of CVs where lines are often strongly in absorption (e.g.\
Hartley \etal\ 2002a,b) or showing P Cyg profiles. We attribute this
to the higher X-ray and EUV irradiation in a magnetic system. 
Strong, broad Ly\,$\alpha$ absorption can be attributed to the white
dwarf, while its narrow emission core is geocoronal emission which
is not completely subtracted during pipeline reduction. 

In addition the spectrum shows narrow interstellar absorption features
which can be used to estimate the column towards AO Psc (e.g.\ Savage
1991). The equivalent width of the 1528-\AA\ Si\,{\sc ii} absorption
line is 0.159~\AA, which, using an oscillator strength 
of $f=0.2303$ (Morton 1991) implies a value of $N_{\rm
Si}=2.5{\times}10^{14}$\,cm$^{-2}$. This suggests (e.g.\ Gnacinski 2003)
that  $N_{\rm H}$\,\sqiglt $10^{20}$\,cm$^{-2}$ and thus that 
\mbox{$E(B-V)$}\,\sqiglt 0.02.

\begin{figure*}       
\includegraphics[angle=-90,scale=0.73]{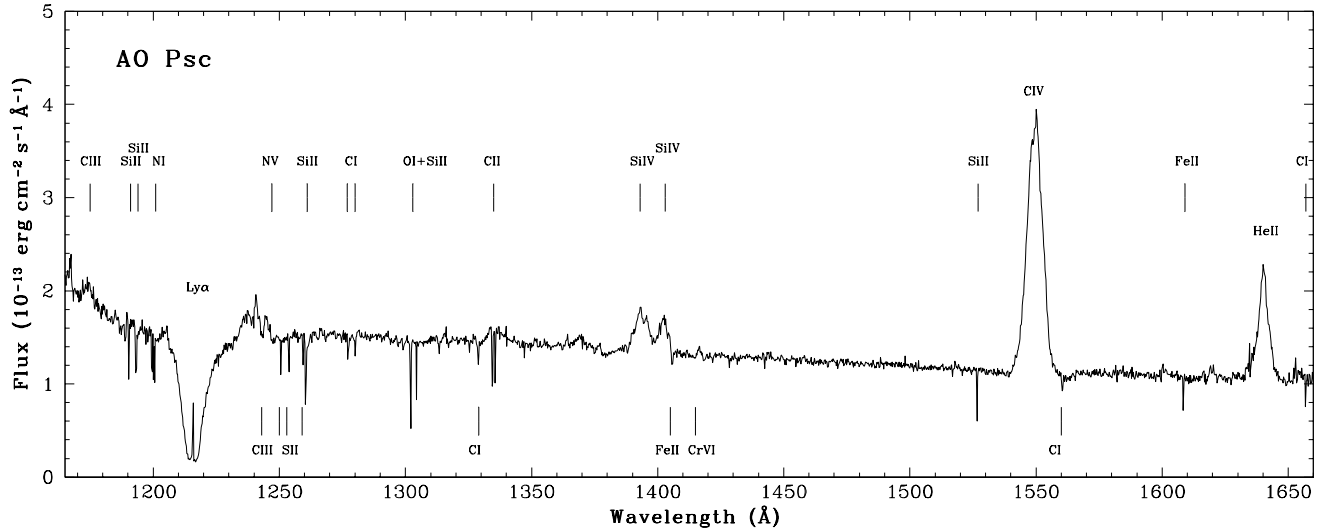}
\caption{The summed {\sl HST}-STIS spectrum of AO~Psc.}
\end{figure*}

\begin{figure*}       
\begin{center}\includegraphics[angle=-90,scale=0.85]{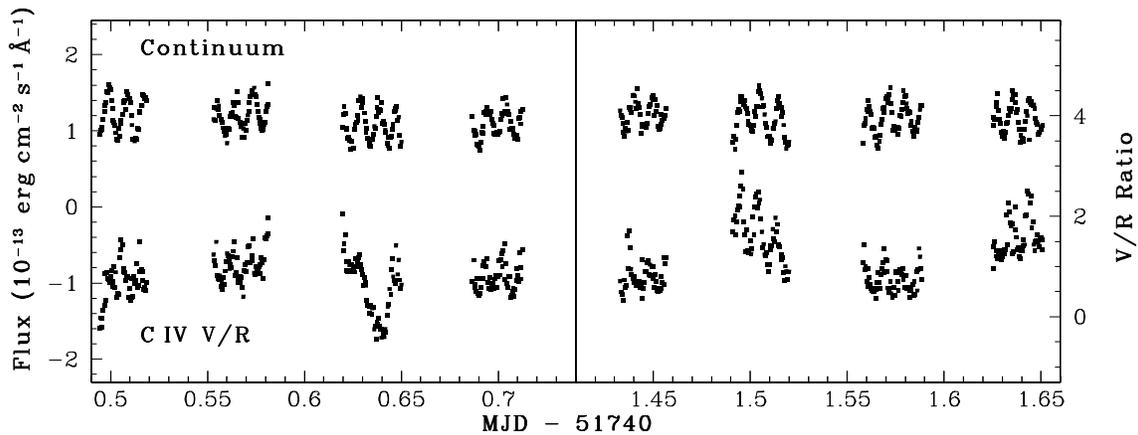}\end{center}
\caption{The continuum flux and the C\,{\sc iv} V/R ratio in each
``exposure''. Variations at the 805-s spin cycle dominant the data.}
\end{figure*}

The continuum and line fluxes, plus the motion of the lines, all show
a prominent variation with the spin cycle (see Fig.~2), as expected
in an intermediate polar.  Fourier analysis of these quantities 
(not shown) finds no beat-cycle periodicity or other periodicities 
shorter than the orbital cycle.   

The focus of this paper is the behavior over the orbital period.  To
reveal this we removed the spin-cycle pulsation by normalizing each
spectrum to the continuum. This process removes most of the
pulsation, but is not perfect, resulting in residual
spin-cycle ``rippling''.

The observations, broken up by {\sl Hubble\/}'s low-Earth orbit,
consist of two ``visits'', each with 4 sections of data $\approx$\,0.2
orbital cycles long, separated by gaps of $\approx$\,0.3 cycles.  It
so happens that these sections interleave to produce full coverage of
AO~Psc's orbit.  We thus display the data against orbital phase as
though they were taken in a continuous sequence (though the caveat
that they were not should be borne in mind).

Fig.~3 shows the region of spectra from 1230 \AA\ to 1415 \AA,
which contains the weaker lines, and Fig.~4 shows the strongest line,
\c4\ (the color scales are not the same in the two plots).
The lines are purely in emission for half an orbital cycle, but the
weaker lines are cut into by prominent absorption features for the 
other half.  \civ\ is in emission except for blue-shifted absorption
around phase 0.73.  To allow the reader to judge the depth of the
absorption we show in Fig.~5 the data from six orbital phases as 
line profiles. 

The emission moves in velocity, being bluest at phase \appro 0.6. The
absorption has the same overall velocity trend as the emission, being
reddest at phase \appro 0.1, but has additional, rapid velocity
variations. We should note that, with only \appro 1.5 orbital cycles
of data, we cannot be certain from this data alone that the absorption
repeats with the orbital cycle.  However, the behavior is similar to
that of the optical Balmer and He\,{\sc i}\ lines, and of the X-ray
continuum, which all show absorption dips occurring at these orbital
phases (H91).

Inspection of Fig.~2 (for example near 1330\AA\ around orbital phase
0.8, and near 1400\AA\ around orbital phase 0) suggests that the 
absorption also varies over the spin cycle, in that it is often 
more pronounced at the minima of the spin pulsation. However, this
is hard to prove conclusively since it is a transient effect and,
as can be deduced from Fig.~2, all measures of the line are dominated
by a spin-cycle variation.

\section{Wind features?}
Absorption features owing to winds are common in the UV lines of 
cataclysmic variables, and the feature seen at phase 0.73 in
\c4\ looks like the classic P~Cygni signature.  However, we consider 
that there are reasons to develop a non-wind model for the 
absorption features in this star. 

(1) Most models suggest that winds are driven from a boundary layer or
inner disk. In an intermediate polar such as AO~Psc there is no
boundary layer or inner disk (an 805-s rotator in equilibrium would
have an inner disk disrupted out to $\approx$\,2\pten{8}\ m).

(2) The absorption in the weaker lines does not give typical P~Cygni
profiles. Instead the absorption is often in the line core, though
there is a bias to the blue.

(3) The absorption troughs have a lower velocity (out to
$\approx$\,--1000 \kmps\ in the weaker lines, --1300 \kmps\ in \civ)
than usual in wind-formed P~Cygni lines.  For example --5000 \kmps\ is
observed in BZ~Cam (Prinja \etal\ 2000), and maximum velocities are
often comparable to the white-dwarf escape velocity.

\begin{figure*}         
\includegraphics[scale=0.70]{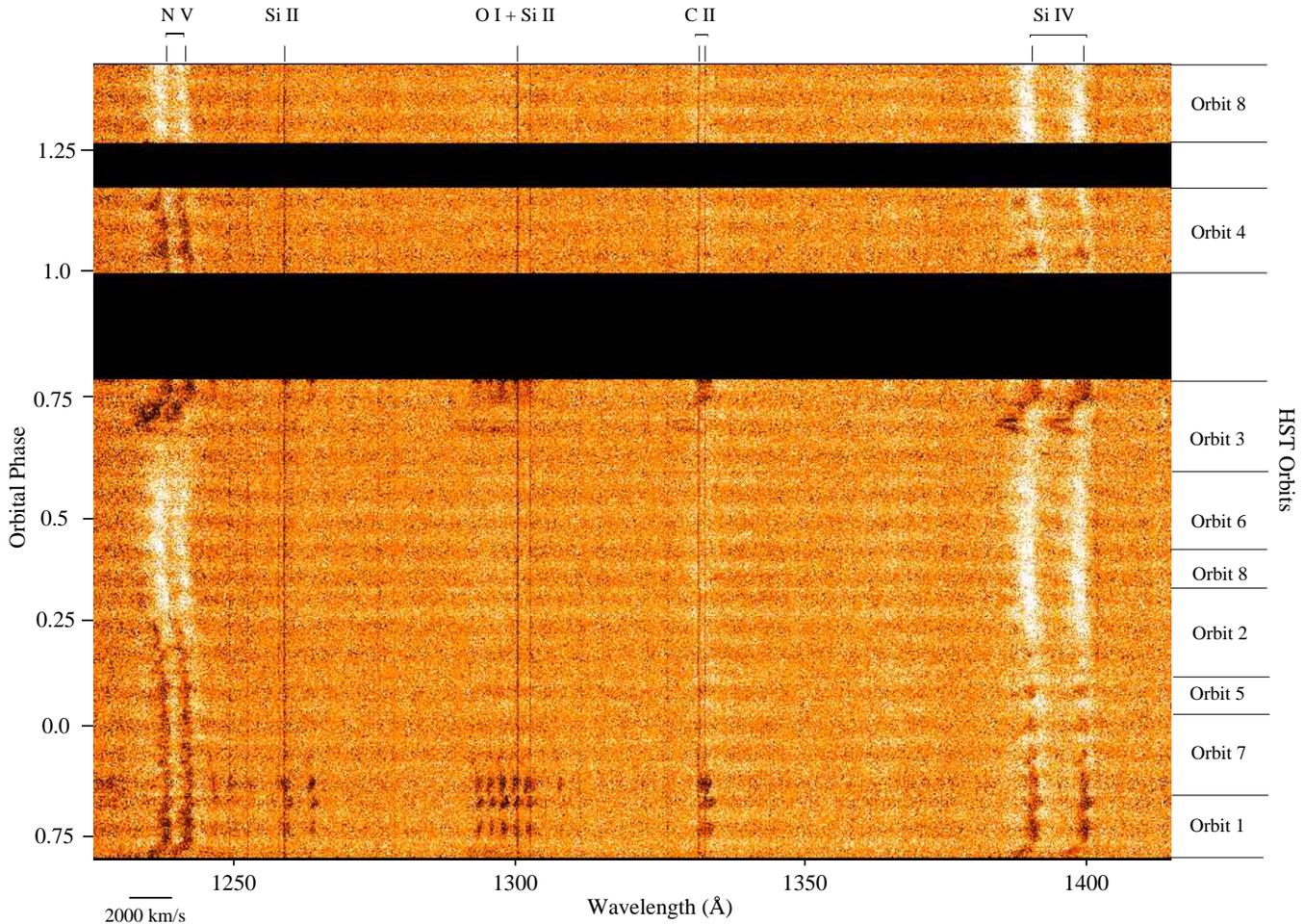}
\caption{Variations of the UV spectrum over orbital phase.  The 
data segments have been interleaved but no binning has been done.
The chosen phase zero is our estimate of inferior conjunction of
the secondary, based on our interpretation of the lines and a 
comparison with the data from H91.  Phase zero is at JD(TDB) 2451740.8920.
An indication of the velocity scale is shown at bottom left. The narrow, 
stationary lines are interstellar. }
\end{figure*}

\begin{figure}         
\includegraphics[scale=0.85]{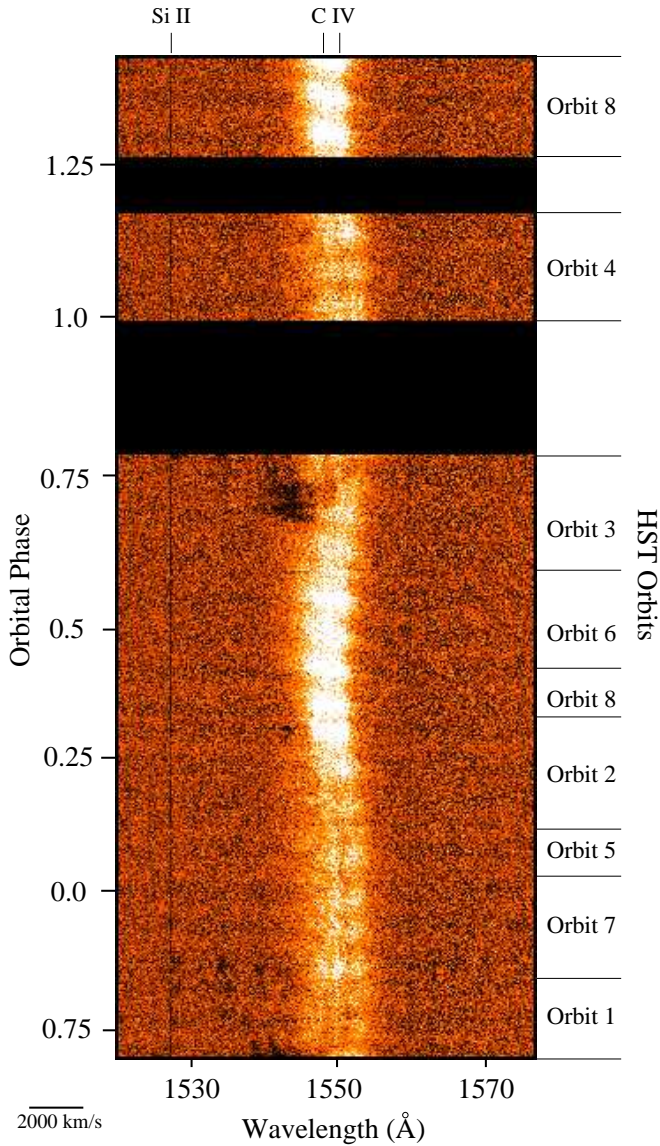}
\caption{As for Fig.~3 but for the region showing \civ\,$\lambda$1550.}
\end{figure}

\begin{figure*}         
\begin{center}\includegraphics[scale=0.85,angle=-90]{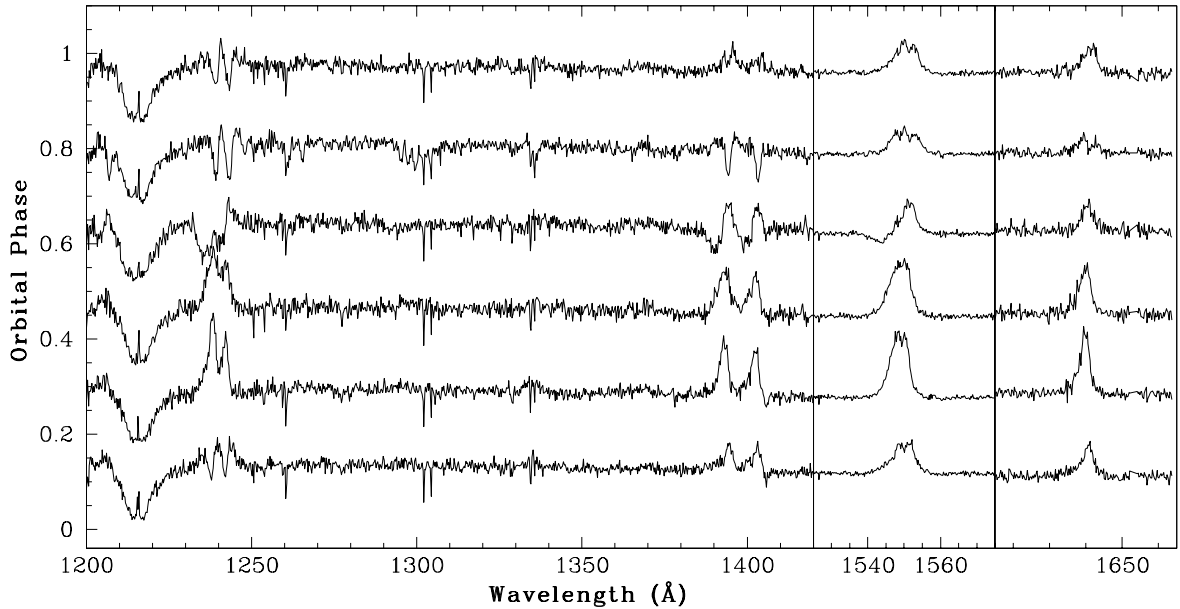}\end{center}
\caption{Line profiles at six orbital phases, showing the variations 
in absorption.  To show both the weak lines and the stronger \civ\ and
\heii\ the three panels have different scales. See Fig.~1 for the
relative intensity of the lines.}  
\end{figure*}

\begin{figure}         
\includegraphics[scale=0.50]{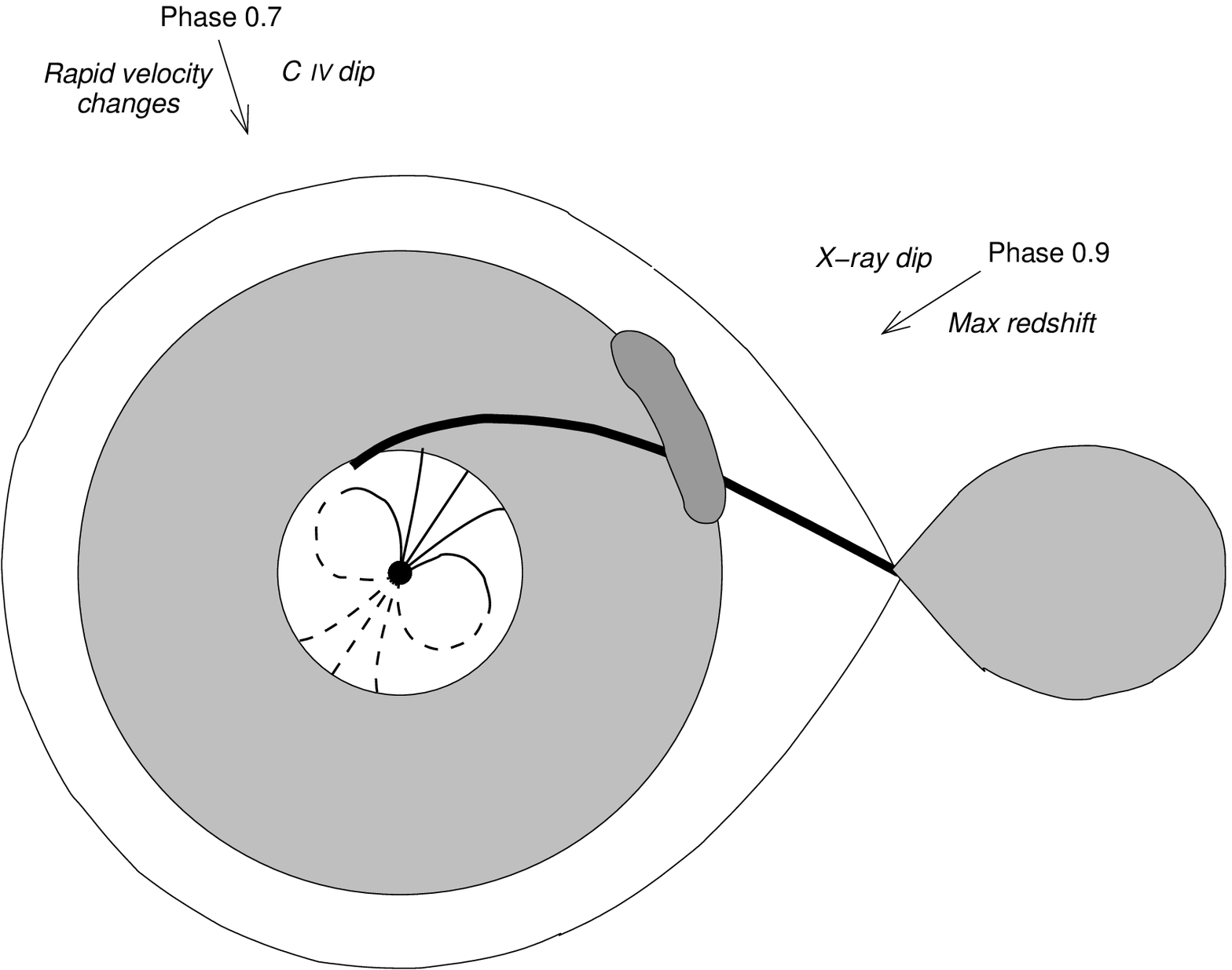}
\caption{Schematic illustration of an overflowing stream in
AO~Psc. The absorption features are seen between phases 0.65
and 0.15.  We label the approximate viewing angle during the
X-ray dip, the \civ\ dip, and during the rapid velocity 
excursions.}
\end{figure}

(4) The features are generally narrower and show more pronounced
variability than in wind-lined systems (for example, compare with the
data shown in Prinja \etal\ 2000; Hartley \etal\ 2002a,b, and Witherick
\etal\ 2003; note that many of these papers display the data after
subtracting the mean profile to enhance the variations, which we have
not done here).  There have been reports of narrow, rapidly varying
wind-formed lines, which are attributed to clumps in an outflow (e.g.\
Hartley \etal\ 2002a), but these are components of broader, deeper
P~Cygni troughs and thus are unlike those in AO~Psc.

It is also worth remarking that the absorption in AO~Psc is markedly 
dependent on orbital phase, being absent for half the cycle.  
Any deduction from this is unclear, however, since, while there is no
particular reason to expect wind-formed lines to depend on orbital
phase, and usually they don't (e.g.\ BZ Cam; Prinja \etal\ 2000),
there are systems where this is indeed seen (e.g.\ V592~Cas; 
Prinja \etal\ 2004) for reasons that aren't understood. 

\section{Stream--field interactions?}  
The fact that the absorption features are so dependent on orbital
phase is a strong indicator that they originate in the accretion
stream. This is supported by the presence of X-ray absorption dips
lasting for \appro 0.2 in phase, which are coincident with absorption
dips in the cores of the optical lines (H91).  Such dips are well
known from X-ray binaries (e.g.\ White 1989) and are often seen in IPs
(Hellier, Garlick \&\ Mason 1993); they are attributed to obscuration
by disk bulges created where the stream hits the disk.

We thus outline an interpretation of the UV absorption features as
originating in the stream. We note that H91 concluded that the motion
of the optical lines was dominated by the stream (rather than by disk
motions). Thus maximum redshift will occur when the stream is flowing
away, near orbital phase 0.9.  This coincides with the optical and
X-ray dips.

With this phasing the UV absorption features occur at phases
0.65--1.15.  We suggest that they arise from the fraction of the
accretion stream that overflows the disk--stream impact.  In the
standard picture of an IP the magnetic field is squeezed in the
orbital plane by the dense accretion disk. However, less dense
material from stream-overflow will encounter field lines further out
than the disk-disruption radius. The material would have to cross
field lines to follow a ballistic trajectory, and so would experience
a magnetic drag, as discussed in the diamagnetic blob model of Wynn
\&\ King (1995).

This interaction would explain why the absorption varies with the spin
cycle. In the standard model for AO~Psc's spin pulse, 
minimum flux occurs when the upper magnetic pole points towards us
(H91). It will present a magnetic barrier to the 
inward flow of the stream lasting for approximately a quarter of a
spin cycle, hence causing an accumulation of material until the
dipole has turned, allowing the material to flow on.  If so, this 
might explain why the absorption increases in the pulse minimum. 

Can we also explain the rapid velocity shifts seen in Fig.~3, for
example at phase 0.70 where the Si\,{\sc iv} line moves 1000 \kmps\ to
the blue and then back redwards on a timescale much faster than the
orbit, but comparable to the spin cycle?  Well, a clue is that it
occurs at the specific orbital phase when the stream--field interaction
region is likely to be viewed in front of the white dwarf (see Fig.~6).  
An ``accretion gating'' scenario in which the material is either pushed
outward, or allowed to flow inward, as the magnetic pole sweeps across 
the region, would produce motion along the line of sight and hence
the rapid changes in Doppler shift seen.  The restriction to one part
of the orbit, and the sudden onset of the features at phase 0.7, 
would then be explained if we only see the absorption features when this
region is in front of the white dwarf, which acts as a UV backlight,

We should also consider the tendancy of the absorption to be
blueshifted in that while it is often close to the rest wavelength it
is typically \sqig 100 \kmps\ bluer than the emission centroid.  Note
that outside the corotation radius the field will be moving faster
than the stream material, and will thus act as a propeller, tending to
slow down the material's infall (which would result in a relative
blueshift for the half-orbit during which we see absorption, when the
stream is on the near side of the white dwarf).  Also, both the
field's kick and the initial impact with the disk could tend to push
the material out of the plane, adding to the blueshift.

Given the propeller effect of the field, it is unclear whether the
overflowing material ends up accreting, or whether it is slowed down,
and perhaps pushed outwards to some extent, so that it ends up merging
with the disk (with an 805-s spin period, any propeller would not
be powerful enough to expell material from the system, unlike
that proposed for faster rotators such as the 33-s system AE~Aqr).  

If material does accrete while remaining confined in orbital
phase it should produce an X-ray modulation at the beat frequency
between the orbital and spin periods, since the accretion geometry
changes on that cycle (e.g.\ Hellier 1991; Wynn \&\ King 1992).
Beat-cycle modulations are not a usual feature of AO~Psc's X-ray
lightcurves, but they have been reported in {\sl EXOSAT\/} data
(Hellier 1991). This suggests that a small amount of the overflow
might be accreting on some occassions, but the dominance of the spin
pulse in the X-ray lightcurves indicates that most of the accretion
flows through the disk.

\section{Comparison with SW~Sex stars}
We have presented evidence that the accretion stream overflows the
accretion disk in AO~Psc, based on an interpretation of absorption
features in the UV lines.  The idea of an overflowing stream was
originated as far back as Lubow \&\ Shu (1976, see also Lubow 1989).
It has since been adopted to explain observational features of several
types of CV. For example, many IPs show the X-ray beat periods that
are characteristic of disk-overflow accretion (e.g.\ Hellier 1991;
Beardmore \etal\ 1998).  FO~Aqr not only shows intermittent beat
periods --- suggesting that overflow is a variable phenomena --- but
also optical absorption features from the overflowing stream (Hellier
\etal\ 1990) that are similar to the UV features reported here in
AO~Psc.

Further, in many non-magnetic systems, eclipse profiles reveal
evidence of a bright stream overflowing the disk (e.g.\ Baptista
\etal\ 1998).  However, stream overflow has been most
widely discussed over its role in the SW~Sex phenomenon, a set of
observational characteristics given a name by Thorstensen \etal\
(1991). Early on, Shafter, Hessman \&\ Zhang (1988) proposed that
overflow was occurring in one of these stars, while Hellier \&\
Robinson (1994) proposed that overflow was the cause of the main
SW~Sex characteristics, namely ``phase 0.5'' absorption features from
the stream as it flowed over the disk, and high-velocity line wings
from the re-impact of the stream with the disk.  Knigge \etal\ (2004)
have shown that phase-0.5 absorption also occurs in the UV lines of at
least one SW~Sex star, while Hynes \etal\ (2001) found similar SW~Sex
characteristics in an LMXB, suggesting that overflow can also occur in
disks around neutron stars.

We can thus compare AO~Psc to SW~Sex stars and conclude that 
stream overflow is likely occurring in both.  Indeed, it is becoming
clear that overflow is widespread in many types of cataclysmic variables,
which means that ``SW Sex star'' becomes less a well-defined class of star
and more a set of observational characteristics resulting (1) from overflow,
(2) from high accretion rate (which appears to be a feature of classic
SW~Sex stars), and (3) from being viewed at high inclination (see,
e.g., Knigge \etal\ 2000). 

If, however, the UV absorption features seen in AO~Psc have a similar
origin to the ``phase 0.5'' absorption characteristic of SW~Sex
stars, why do they have virtually the opposite phasing, being seen
at phases 0.65--1.25 in AO~Psc, but at pases 0.2--0.6 in SW~Sex stars?
Our explanation is as follows. 

The conventional SW~Sex stars are high-inclination eclipsing systems
in which the white dwarf is usually hidden by the optically dim wall
of a flared disk. We thus see bright disk surface only beyond the 
white dwarf.  In the model developed
in Hellier (1996; 1998) the optical absorption features result when
the overflowing stream obscures the optically bright disk surface, 
which is only during phases 0.2--0.6.  

Now, AO~Psc is a lower-inclination, non-eclipsing system where
any disk flare will have a lesser effect.  Further, the UV absorption
features require a UV backlight, which can only be provided by 
regions near the white dwarf, and not by the cooler outer disk. 
These difference could explain the marked change in when 
absporption is visible in AO~Psc, even though its origin is similar 
to that in SW~Sex stars. 

It is noteworthy that, in AO~Psc, SW~Sex characteristics are
occurring in a system which is undoubtably magnetic. This is
important, first, in that it enables us to observe the interaction of
the overflow with the field of the white dwarf, which results in rapid
variability as the stream appears to waggle about under the influence
of the spinning dipole.  Second, it bears on the suggestion by
Rodriguez-Gil \etal\ (2001, see also Patterson \etal\ 2002) that all
SW~Sex stars are magnetic, and that the magnetism is fundamentally
liked to SW~Sex behavior.

If SW~Sex characteristics are largely a consequence of stream
overflow, then this can indeed occur in a magnetic system, but is
unlikely to be caused by the magnetism.  Thus, SW~Sex characteristics
would be merely coincidental to AO~Psc's magnetism.  This would concur
with the fact that many SW~Sex stars are not strong X-ray sources, let
alone showing the X-ray pulsations that are prominent in AO~Psc, and
are thus unlikely to have fields strong enough to control the
accretion flow.

\acknowledgments
The work presented in this paper is based on observations with the 
NASA/ESA Hubble Space Telescope, obtained by the Space Telescope 
Science Institute which is operated by AURA Inc under NASA contract 
No. 5-26555.  We thank a referee for prompting clarifications and
improvements of this work, and Bill Welsh for advice and assistance.

\end{document}